\documentstyle[hip-artc]{article}  
\volnumber{xx}  \edyear{2004}  \frompage{000} \topage{000}               %| 
\recrevdate{xx xx     2004}                                              %| 
\def\rightmark{Balance functions} 
\def\leftmark{P. Bo\.zek et al.}
 
\title{Balance functions from a thermal model \footnote{\small Talk presented
by WF during the 3rd Budapest Winter School on Heavy-Ions, \\ Dec. 8-11,
2003, Budapest, Hungary. Supported in part by the Polish State \\
Committee for Scientific Research grant 2 P03B 059 25.}} 
\authors{
{\twerm  W. Florkowski$^{1,2}$, P.~Bo\.zek$^1$, and W.~Broniowski$^1$%
}\\[2.812mm]
{\normalsize
\hspace*{-8pt}$^1$ The H. Niewodnicza\'nski Institute of Nuclear Physics, \\ 
Polish Academy of Sciences, PL-31342 Krak\'ow, Poland\\[0.2ex] 
\hspace*{-8pt}$^2$ Institute of Physics, \'Swi\c{e}tokrzyska Academy,\\ 
PL-25406 Kielce, Poland
}}
 
\abstract{A calculation of the pion balance functions in a thermal
model is presented. The total result consists of resonance and
non-resonance parts. A satisfactory agreement with the data on Au+Au
collisions at $\sqrt{s_{NN}}$=130~GeV is found. } 
 
\begin{document}
 
\maketitle

\section{Introduction}

In the last years the thermal (statistical) models of particle
production in ultra-relativistic heavy-ion collisions turned out to be
very successful in describing hadron yields
\cite{pbmrhic,mich,rafQM02}, spectra \cite{wbwfspectra}, the
elliptic-flow coefficients $v_2$ \cite{v2hbt,v2BL}, and the HBT radii
\cite{v2hbt,hbtBL}. In view of this fact it is interesting to study
further observables in the thermal approach and compare the results of
the model calculations with the data. One particular example of such
observables are balance functions \cite{bass,jeon}, which describe
correlation between opposite-charge particles in the rapidity space
and are closely related to charge fluctuations
\cite{jeonkoch,asakawa}.

The balance functions measured by the STAR Collaboration
\cite{STARbal,phdmsu} at RHIC are defined by the formula
\begin{equation}
B(\delta,Y) = {1\over 2} 
\left\{
{\langle N_{+-}(\delta) \rangle - \langle N_{++}(\delta) \rangle \over
\langle N_+ \rangle} 
+
{\langle N_{-+}(\delta) \rangle - \langle N_{--}(\delta) \rangle \over
\langle N_- \rangle} \right\},
\label{def}
\end{equation} 
where $N_{+-}(\delta)$ is the number of the opposite-charge pairs such
that both members of the pair fall into the rapidity window $Y$ and
their relative rapidity is $|y_2-y_1|=\delta$, $N_{+}$ is the number
of positive particles in the interval $Y$, and other quantities are
defined in an analogous way.

The measurement \cite{STARbal} showed that the widths of the balance
functions are smaller than expected from models discussed in
Ref.~\cite{bass} and significantly smaller than observed in elementary
particle collisions.  This problem was discussed by Bialas in
Ref. \cite{Bialas}, where the small widths of the balance functions
were explained in the framework of the coalescence model \cite{coal}.
In this paper we present the results of an alternative calculation of
the $\pi^+ \pi^-$ balance function \cite{ourbal}, which is based on a
thermal model with resonances \cite{wbwfspectra}.  In a more recent
paper by Pratt et al. \cite{many}, it was shown that the measured
balance function may be well reproduced in the canonical blast wave
model.

\begin{figure}[tb]
\begin{center}
\hspace{1.50cm} \epsfysize=6.0cm \epsfxsize=8.0cm \epsfbox{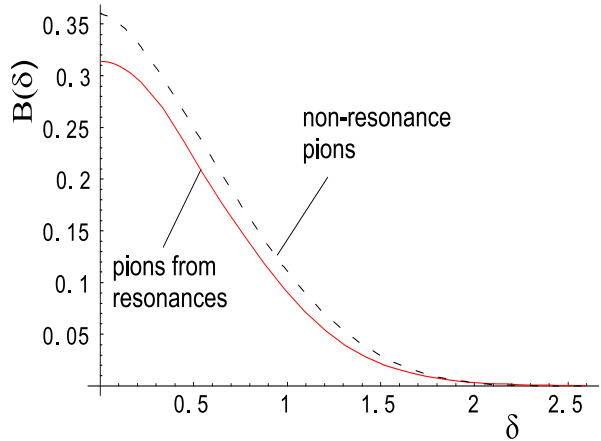}
\end{center}
\vspace{-0.5cm}
\caption{Contributions to the balance function from the neutral
resonances (solid line) and from the non-resonance pions (dashed
line), plotted as a function of the rapidity difference of the two
pions. The temperature of the system is $T$ = 165 MeV, and the
expansion parameters of the model correspond to the average transverse
flow $\langle \beta_\perp \rangle=0.5$.}
\label{brnr}
\vspace{-0.5cm}
\end{figure}

In our approach, the $\pi^+ \pi^-$ balance function has two
contributions related to two different mechanisms of the creation of
an opposite-charge pair. The first one {\it (resonance contribution)}
is determined by the decays of neutral hadronic resonances, whereas
the second one {\it (non-resonance contribution)} is related to other
possible correlations among the charged particles. We assume that the
second mechanism forces the two opposite-charge pions to be produced
at the same space-time point with thermal velocities. On the other
hand, the first contribution refers only to the decays of {\it
neutral} resonances which have a $\pi^+ \pi^-$ pair in the final state
(we explicitly include $K_S, \;\eta, \; \eta^\prime, \; \rho^0, \;
\omega, \; \sigma, \; {\rm and} \; f_0$). In this case the
correlations among emitted pions are completely determined by the
kinematics of the decays. The details of the modeling of the
non-resonance contribution and other technical remarks are given in
Ref. \cite{ourbal}.

\begin{figure}[h]
\begin{center}
\hspace{0.5cm} \epsfysize=8.0cm \epsfxsize=10.0cm \epsfbox{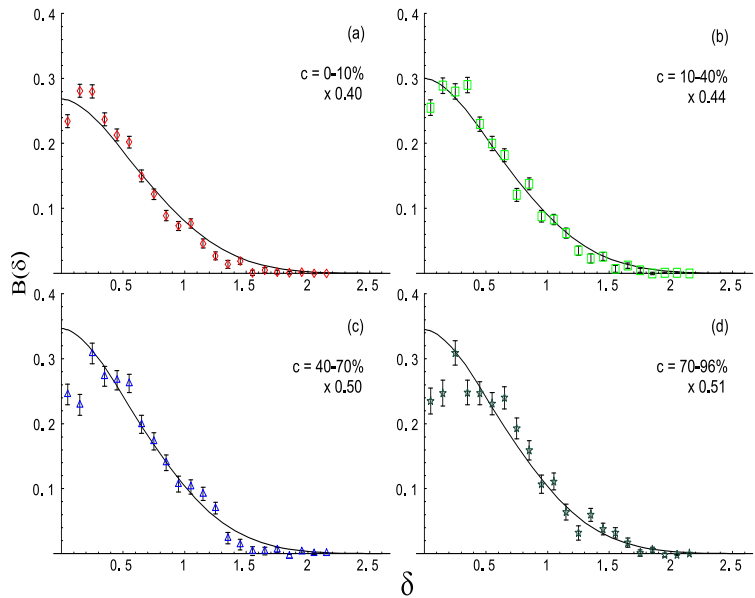}
\end{center}
\caption{Balance functions for the pions in the thermal model
calculated for four different centrality classes and compared to the
experimental data of Ref.~\cite{STARbal}. The normalization of the
model curves was adjusted in each case and is listed near the plot
labels.}
\label{thdat}
\vspace{-0.5cm}
\end{figure}

\section{Results}

According to the discussion presented above, the $\pi^+ \pi^-$
balance function can be constructed as a sum of the two terms
\begin{equation}
B(\delta) = B_{\rm R}(\delta) + B_{\rm NR}(\delta).
\label{balsum}
\end{equation}
The functions $B_{\rm R}(\delta)$ and $B_{\rm NR}(\delta)$ resulting
from our model calculation are presented separately in Fig. 1. The
value of the temperature used in the calculation was $T$ = 165 MeV,
and the expansion parameters of the model were fitted to the spectra
of hadrons. This procedure yields the average transverse flow of
0.5$c$. One can observe that the widths of the two contributions are
similar. The calculated total width, $\langle \delta \rangle = 0.66 $,
turns out to be slightly larger than the experimental value for the
most central collisions. The STAR result for the most central
events ($c=0-10\%$) is $\langle \delta \rangle = 0.594\pm 0.019$, for
the mid-central ($c=10-40\%$) $\langle \delta \rangle = 0.622\pm
0.020$, for the mid-periferal ($c=40-70\%$) $\langle \delta \rangle =
0.633\pm 0.024$, and for the periferal ($c=70-96\%$) $\langle \delta
\rangle = 0.664\pm.029$. Such dependence of the width of the balance
function on centrality cannot be reproduced in our model by changes of
the transverse flow within limits consistent with the single-particle
spectra.

In Fig. 2 our results are compared with the experimental values
obtained for four different centrality classes. The normalization of
the model curves was adjusted in each case, since we were not able to
take into account, in a different way, the effect of a limited
detector efficiency and acceptance. On the other hand, the kinematic
cuts in pseudorapidity and transverse momentum were included exactly
in \cite{ourbal}. The shapes of the model balance functions agree well
with the data except for the most central case where the theoretical
width is slightly larger. We note that dips of the experimental
balance functions at very small values of $\delta$ are caused by the
HBT correlations. This type of effects is not included in our
approach.  We also note that the effects of the detector efficiency
may influence the width of the balance function.

As a conclusion we state that our simple thermal model with thermal
and expansion parameters fixed earlier by fitting the ratios of
particle abundances and the transverse-momentum spectra gives a
quite satisfactory description of the balance functions.

\vfill\eject
\end{document}